# E-coherent crystalline interfaces: coherency enhanced by discohesion arrays

Ryan B. Sills[1*], Alejandro Hinojos[2], Trevor J. Murray[1], Shane H. Cooley[1], Xiaowang W. Zhou[2], and Douglas L. Medlin[2]

[1]Department of Materials Science and Engineering, Rutgers University, Piscataway, NJ 08854

[2]Sandia National Laboratories, Livermore, CA 94551

**Abstract**

Coherent crystalline interfaces form when a pair of joined crystals share lattice sites. Such interfaces are ubiquitous in materials, minerals, and compounds, with examples including grain boundaries in polycrystals and phase boundaries in multi-phase systems. Existing methodologies such as the topological model[1] provide a framework for understanding the nature of coherency between two crystals and the line defect content within an interface. However, these methods only consider states of coherency achieved via affine transformations. Here we show that in some interfaces, local relaxations in the form of non-affine transformations lead to the introduction of additional coincidence sites within the interface; we term this class of interfaces as *e-coherent*. These non-affine relaxations are topologically equivalent to inserting disconnection (or disclination) dipoles or loops into the interface. Unlike traditional interfacial line defects, the defects associated with e-coherency cannot have long-range stress fields and their motion alters the state of coherency between the crystals. Given these unique properties, we differentiate them from other defects by referring to them as *discohesions*. Through atomistic simulations and transmission electron microscopy, we show that the energetics and kinetics of e-coherent interfaces are strongly affected by the discohesion content in the interface, leading to fundamentally different behaviors compared to non-e-coherent interfaces. We demonstrate e-coherency in grain, twin, and phase boundaries, and that a given interface can have multiple possible e-coherent states. These results suggest that e-coherency is likely to be pervasive in crystalline solids.

**Introduction**

The atomic structure of interfaces in crystalline solids strongly influences response to mechanical[2], thermal[3], and radiation stimuli[4]. Interfaces often possess line defects such as dislocations, disconnections, and disclinations[5], which affect the thermodynamics, kinetics, and interactions of the interface[6–9]. In traditional analyses of coherent interfaces[10], coherency between an adjacent pair of crystals $A$ and $B$ is established by applying affine transformations. These transformations are in the form of deformation gradients $^A\mathbf{F}$ and $^B\mathbf{F}$ which impose rotations and/or stretches/compressions that result in (affine) coincidence site (CIS) atoms in the interface, as shown in Fig. 1. Here, we define CIS atoms as atoms which are simultaneously part of both crystals. Once coherency is established, the dichromatic pattern/complex and the topological model can be utilized to define possible line defects—dislocations and disconnections—that are admissible within the interface[1,10]. (Disconnections are line defects, specific to interfaces, which possess both step and dislocation character[11].)

In addition to the introduction of line defects, the atomic structure of interfaces is often affected by local atomic relaxations which act to lower the energy of the interface. Concepts such as the structural unit model[12,13], polyhedral unit model[14,15], and grain boundary complexions/phases[16–18] provide powerful tools for describing how these atomic relaxations alter the structure of interfaces. In many situations, addition or removal of atoms near the interface has a strong influence on the relaxations[19] and may lead to the formation of new structural units/phases[16,20,21]. Another concept related to atomic relaxations is the notion of atomic shuffling during interface migration, whereby locally varying atomic displacements are necessary to accommodate migration[22]. These displacements can be thought of as local relaxations that occur during migration. Finally, we note that in the case of interfaces with misoriented and/or misfitting crystals, additional long-range relaxations may occur upon insertion of dislocations and/or disconnections within the interface which accommodate the misorientation and/or misfit[10]; This gives rise to so-called semi-coherent interfaces.

In this work we show that in some interfaces, these local atomic relaxations and possible addition/removal of atoms can be thought of as non-affine transformations which lead to additional atoms being shared between the crystals, as shown schematically in Fig. 1. In the relaxed interface, these additional non-affine CIS atoms are indistinguishable from traditional affine CIS atoms (as shown by our results below); both sets of CIS atoms are shared by the crystals, with some local distortions due to the non-affine relaxations. This type of interfacial relaxation, which is only exhibited by some interfaces, therefore leads to an enhanced state of coherency between the crystals, and we term such interfaces as *e-coherent* (short for enhanced-coherent). Many important properties of interfaces are affected by the nature of interfacial coherency, and we demonstrate below how e-coherent interfaces behave differently than traditional coherent interfaces. This concept also provides a powerful new approach for interrogating the atomic-scale dynamics of interfaces since the line defects associated with e-coherency, which we term *discohesions*, can be rigorously tracked in computational simulations along with more conventional line defects.

**Results**

Closely examining the non-affine relaxations (which may include addition/removal of atoms) that give rise to e-coherency reveals that they are equivalent to insertion of line defects in the form of disconnection (or disclination) dipoles or loops, as shown in Fig. 1. In comparison to traditional dislocations and disconnections, these line defects exhibit two unique properties: (1)

they can only exist in the form of dipoles and/or loops and (2) they introduce additional (non-affine) CIS atoms into the interface. Given these unique properties and that these defects only exist in some interfaces, we argue that they are worth differentiating from other defects. As such, we term them as *discohesions*, given their role in altering the state of coherency. The fact that discohesions lack long-range fields means that they cannot relax misorientations and/or misfits, indicating that e-coherency and semi-coherency are independent concepts. Similar to the topological model[1], Fig. 1 shows that a discohesion can be specified by the vectors $^A\boldsymbol{t}_1$, $^B\boldsymbol{t}_1$, $^A\boldsymbol{t}_2$, and $^B\boldsymbol{t}_2$ which define its Burgers vector (below $i$ implies that either set of vectors 1 or 2 could be used)

$$\mathbf{b} = {^A\boldsymbol{t}_i} - {^B\boldsymbol{t}_i}, \tag{1}$$

and its step height

$$h = \hat{n} \cdot ({^A\boldsymbol{t}_i} + {^B\boldsymbol{t}_i})/2 \tag{2}$$

where $\hat{n}$ is the terrace plane unit normal vector, and its spacing (for straight discohesions)

$$s = \left| (\hat{\xi} \times \hat{n}) \cdot ({^I\boldsymbol{t}_1} - {^I\boldsymbol{t}_2}) \right| \tag{3}$$

where $\hat{\xi}$ is the sense vector of the discohesion (pointing in the direction of the discohesion line) and $I$ denotes crystal $A$ or $B$ (in the case of loop discohesions, the diameter can be defined similarly, see Supplemental Information). Note that we define a discohesion here as both disconnections in the dipole for straight discohesions or the entire loop for loop discohesions (since they can only exist as dipoles and loops).

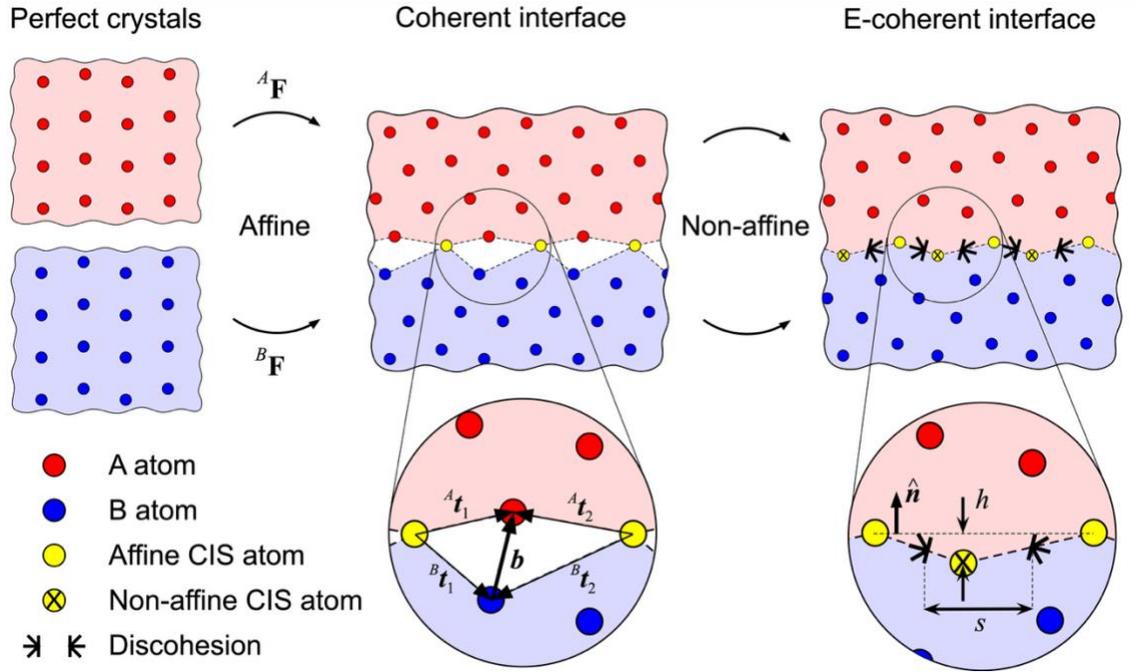

*Fig. 1. Schematic depiction of the construction and geometry of symmetric-tilt coherent and e-coherent interfaces. Discohesions are inserted by locally distorting each crystal to render atom pairs (one from each crystal) coincident (and removing one of the atoms in each pair). Each marker for the discohesion denotes a disconnection, since discohesions are composed of disconnection dipoles (or loops). The arrowheads in the markers point to the centroid of the discohesion. CIS = co-incidence site*

Through molecular dynamics (MD) simulations and high-resolution scanning transmission electron microscopy (HR-STEM) (see Methods), we have identified e-coherent twin, phase, and grain boundaries, and have explored their behavior and properties. The interfacial line defect analysis (ILDA) method for characterization of atomic simulation data and the 2D polyhedral template matching (2D-PTM) method for HR-STEM image characterization previously developed by the authors[23,24] enable us to perform a detailed assessment of atomic and line defect dynamics in e-coherent interfaces. In particular, ILDA automatically identifies interfacial line defects in an MD simulation through construction of Burgers circuits across the interface using CIS atoms.

These CIS atoms are rigorously identified using polyhedral template matching (PTM) which compares each atom's local environment to atomic templates from perfect crystal structures[25]; utilizing PTM in our analysis safeguards against overinterpretation of extensive atomic relaxations near the interface that cannot be characterized in terms of line defect insertion events. In our results below, we show that ILDA consistently identifies discohesions in e-coherent interfaces, which appear as disconnection dipoles and loops (since ILDA is designed to identify disconnections). In traditional coherent interfaces (which are not e-coherent), no line defects are observed. This comparison points to the physical significance of the discohesion concept and supports our interpretation of e-coherency in terms of insertion of discohesions defined in a manner consistent with that of other line defects.

As an example of an e-coherent interface, the middle of Fig. 2(a) shows an HR-STEM atomic resolution image of a $\{1\,0\,\bar{1}\,1\}$ twin observed in the hexagonal close-packed (HCP) ε-martensite phase in stainless steel. This structure is consistent with our MD simulations (overlaid on the image) and is equivalent to the structures reported previously for $\{1\,0\,\bar{1}\,1\}$ twins in other HCP metals[26–30]. Note that the atoms at the interface are aligned into a single planar layer, whereas away from the interface in the single crystals the $\{1\,0\,\bar{1}\,1\}$ planes are corrugated (e.g., atoms zig-zag up and down). Because of this corrugation, were the interface to be formed purely by an affine transformation, only every other atom near the interface would be coincident, as is shown in the top of Fig, 2(a). In real systems, however, atoms relax, and in this specific interface, this relaxation gives an enhanced state of coherency such that every atom the interface plane is a CIS atom (CIS atoms identified by ILDA and PTM are colored yellow in Fig. 2).

It is this local relaxation to an enhanced state of coherency that is quantitatively described in our theory by discohesions. Specifically, to render all atoms coincident within the interface plane is accomplished by inserting discohesion arrays with Burgers vector $\mathbf{b}_{coh1}$ = [0.0 0.0 -0.65] Å, a step height of $h = 0$, and a spacing of $s = 2.36$ Å into the interface, as is shown by the array of line defects revealed by ILDA in the bottom of Fig. 2(a) (each line defect is half of the dipole comprising a discohesion). In MD we can instead construct the twin interface based on the affine state of coherency with both atom columns near the interface present, shown in the top row of Fig. 2(a). Doing so results in an unstable interface (which quickly disorders) because the atom pairs are only 0.65 Å apart, leading to high energy. By removing one column of atoms from each pair, the system is able to relax into the stable e-coherent structure.

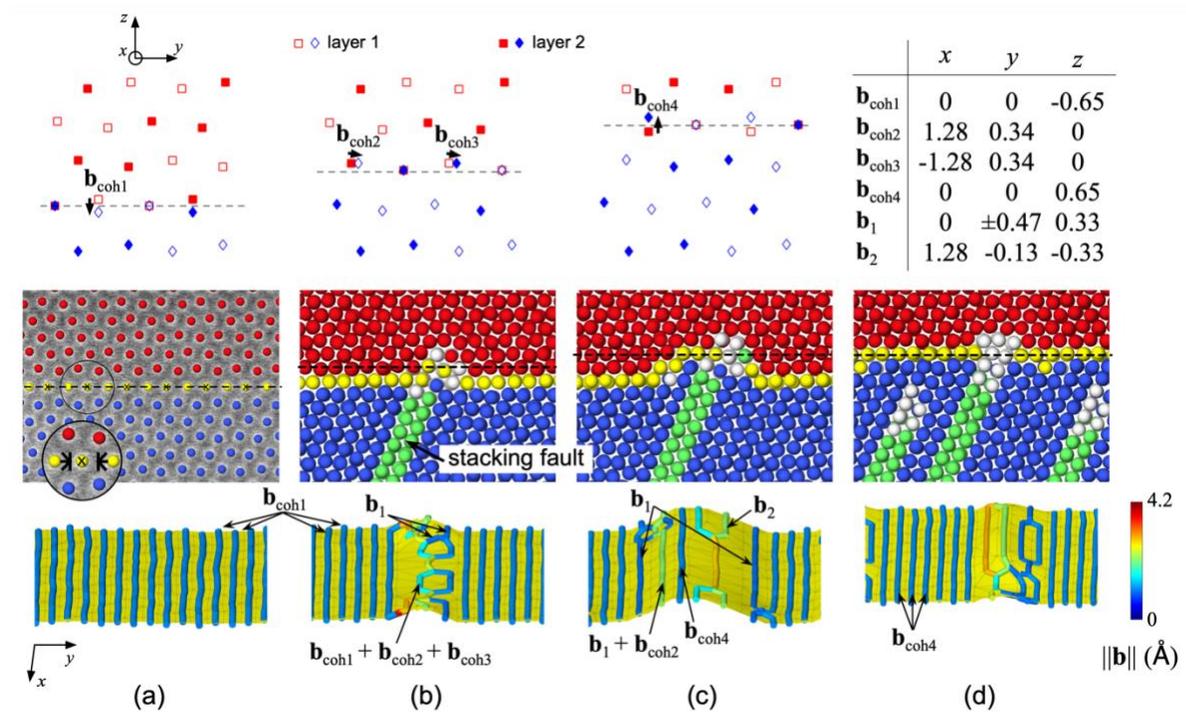

Fig. 2 Migration of an e-coherent $\{1\,0\,\bar{1}\,1\}$ HCP twin. Top row: Atomic structure of coherent interface (layers 1 and 2 refer to successive planes of atoms along the x-direction). Middle row: MD simulation snapshots. Bottom row: line defects identified by ILDA. (a) Initial geometry, with the middle image showing comparison between MD and HR-STEM. Blue lines in the MD snapshot are the discohesions in the pristine $\{1\,0\,\bar{1}\,1\}$ twin. (b) A disconnection nucleates, causing motion

*upward by one plane. (c) Another disconnection nucleates, causing motion upward by an additional plane. (d) Both disconnections migrate across the interface. The table at the upper right gives the components of the Burgers vectors in units of Å.*

To explore the influence of e-coherency on migration of the interface, we apply shear strain to the system. The sequence of images (Figures 2b-d) shows the migration of the twin boundary upwards by two atomic planes, with the top row of images shows the atomic structure with affine transformations (rotations) applied, the middle row shows the atomic structure from MD, and the bottom row shows the interface plane with line defects from MD identified by ILDA. A partial dislocation first nucleates at the free surface of the bottom crystal (not visible) and impinges on the interface, as indicated by the stacking fault in Fig. 2(b). This interaction then stimulates the nucleation of a disconnection loop with Burgers vector $\mathbf{b}_1$ and a step height of one atomic plane which begins propagating the interface upwards. However, this process is messy and disordered with the interface becoming choppy and topologically complex. Furthermore, rather than propagating the disconnection along the interface to migrate it upwards by one atomic plane, another disconnection loop quickly nucleates with Burgers vector $\mathbf{b}_2$ and a step height of one plane to translate the interface upwards by an additional plane (see Fig. 2(c)). This $\mathbf{b}_2$ disconnection then propagates along the interface to bring the entire interface upward (see Fig. 2(d)) by two planes. From there, this process repeats again, leading to propagation upwards by an additional two planes (not shown).

Disconnections for $\{1\ 0\ \bar{1}\ 1\}$ twins have been extensively analyzed theoretically[26,27,29–31] and experimentally[32–34], concluding that one-step disconnections are not observed (unless there are additional reactions in the interface[29,32], see Supplemental Information) and unfavorable because their propagation requires a large amount of shuffle[27]. Our concept of e-coherency provides deeper insight into the unfavorability of such one-step disconnections; while shuffle describes the dynamics of each disconnection, the differences in e-coherency between one- and two-step disconnections cause the energies of the associated twin boundaries to be different. In the initial plane, discohesions with Burgers vector magnitude $||\mathbf{b}_{coh1}|| = 0.65$ Å occupy the entire interface. In contrast, in the next plane up, the discohesions alternate between Burgers vectors $\mathbf{b}_{coh2}$ and $\mathbf{b}_{coh3}$ both of which have a much larger magnitude of 1.32 Å (primarily due to the large screw component in the x-direction). In dislocation theory[5], line defect energy scales as $||\mathbf{b}||^2$ meaning the interface containing $\mathbf{b}_{coh2}$ and $\mathbf{b}_{coh3}$ will have higher energy than the interface containing $\mathbf{b}_{coh1}$. Propagating upwards one more plane restores a uniform set of discohesions with Burgers vector $\mathbf{b}_{coh4} = -\mathbf{b}_{coh1}$, with comparable energy to the original plane (since $\mathbf{b}_{coh4}$ and $\mathbf{b}_{coh1}$ have the same magnitude). In other words, our theory suggests that single-step disconnections are unfavored due to the associated high twin boundary energy, rather than the large amount of shuffle associated with their motion.

To further illustrate the concept of e-coherency we next consider a different interface in the same steel alloy, an α'–ε interface (α' is the body-centered cubic (BCC) form of martensite). The crystals are oriented with $(1\,1\,0)_{\alpha'} \parallel (1\,0\,\bar{1}\,1)_{\varepsilon}$ and $[1\,\bar{1}\,\bar{1}]_{\alpha'} \parallel [\bar{2}\,1\,1\,0]_{\varepsilon}$. Applying the deformation gradient

$$^{\alpha'}\mathbf{F} = \begin{bmatrix} 1.020 & -0.090 & 0 \\ 0 & 1.002 & 0 \\ 0 & 0 & 1 \end{bmatrix}$$

to the BCC crystal leads to the affine coherent atomic structure shown in the top of Fig. 3(a). This deformation gradient brings half of the atoms near the interface into coincidence, alternating atom columns are not coincident. As with our previous example of the $\{1\,0\,\bar{1}\,1\}$ twin, the structure formed purely by an affine transformation, disagrees with both our HR-STEM and MD simulations (shown in the bottom of Fig. 3(a)) for which all atoms near the interface are coincident (shared by both crystals). This e-coherent interface structure is obtained by inserting discohesions with Burgers vector $\mathbf{b}_{coh1}$ = [0.64 -0.17 0.33] Å, height $h$ = 0.16 Å, and separation distance $s$ = 2.45 Å. As before, this interface is unstable in the absence of discohesions (e.g., if constructed based on affine coherency). In MD we constructed the interface so that all coherency strain is applied to the BCC crystal, providing a driving force for interface migration into the BCC crystal. Two disconnection modes are observed, as shown in Fig. 3(b): disconnection $\mathbf{b}_1$ has magnitude $\|\mathbf{b}_1\|$ = 0.36 Å and leads to the same e-coherency with $\mathbf{b}_{coh1}$, whereas disconnection $\mathbf{b}_2$ has magnitude $\|\mathbf{b}_2\|$ = 0.99 Å and results in e-coherency with oppositely signed discohesions $\mathbf{b}_{coh2}$ = -$\mathbf{b}_{coh1}$.

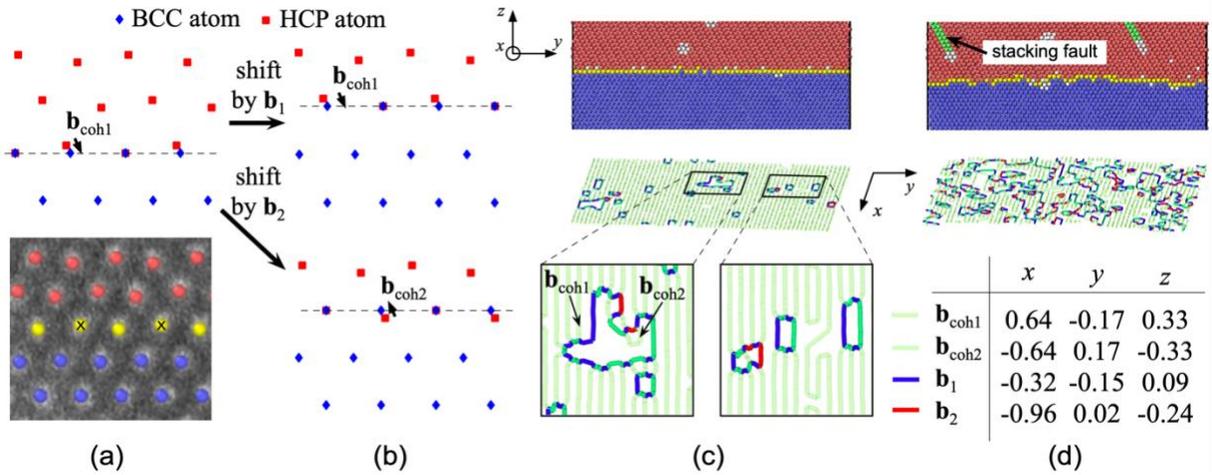

Fig. 3 Migration of an e-coherent BCC-HCP interface. (a) Interface structure showing affine coherent geometry with discohesion $\mathbf{b}_{coh1}$ marked and comparison of e-coherent structure from MD with HR-STEM image. (b) Coherent atomic structure when the interface propagates via disconnections $\mathbf{b}_1$ and $\mathbf{b}_2$. (c-d) MD snapshots of migration into the BCC crystal. (c) Geometry of loops nucleated in the interface. (d) Line geometry after loop expansion and merger. The table at the lower right gives the components of the Burgers vectors in units of Å.

This framework provides insight into how this interface can migrate. Figs. 3(c-d) show MD snapshots analyzed with ILDA revealing the topological mechanisms enabling migration of the interface. Initially the interface is planar and contains a regular array of discohesions (faint green lines). Disconnection loops nucleate all around the interface, with a few examples shown in (c). When the loops are small, they are entirely composed of disconnection mode $\mathbf{b}_1$, though some portions of the loop react locally with discohesions giving a different Burgers vector. Once the loops become larger, the migrated interface plane becomes e-coherent with discohesion $\mathbf{b}_{coh2}$. These discohesions react with the disconnection loop and cause portions of the loop to change from disconnection $\mathbf{b}_1$ (blue lines) to disconnection $\mathbf{b}_2$ (red lines). This line topology is distinct from traditional disconnection loops since the Burgers vector of the loop varies around the loop (while still conserving the Burgers vector). Furthermore, interactions between the loops and the discohesions cause highly irregular loop geometries, which differ from traditional line defect loops that naturally adopt smooth geometries to minimize their energy[5]. These loops continue expanding and reacting with each other, leading to the complex interface line topology shown in Fig. 3(d).

As a final example, we demonstrate e-coherency in a face-centered cubic (FCC) grain boundary (GB). The simulated GB is a $\Sigma 385$ mixed tilt-twist boundary previously studied by Olmsted et al.[37,38]. The coherent atomic structure for the GB is shown in Fig. 4(a). In contrast to the previously considered interfaces, affine CIS atoms are rather far apart in this interface, with many atoms from both crystals in between them. Analyzing the dichromatic pattern reveals three proximate pairs of atoms giving potential discohesion Burgers vectors with magnitudes of 0.58, 0.86, and 1.37 Å. We may expect that the two crystals would prefer to share a single atom rather than having a pair of atoms close together, giving an e-coherent interface. Upon constructing and relaxing the GB, ILDA does in fact reveal discohesion dipole arrays with Burgers vector $b_{coh2}$, separation distance $s = 8.2$ Å, and height $h = 0.69$ Å as shown in Fig. 4(b).

Next, we apply a synthetic driving force to induce interface migration[39]. As the interface begins moving, a set of disconnections causes the interface to transition from e-coherent to coherent with no interfacial line defects. Subsequently, an additional set of disconnections causes the interface to migrate into another e-coherent state featuring $b_{coh3}$ discohesions with the same separation distance and line orientation as the $b_{coh2}$ discohesions, but a step height of $h = 0$. As the interface continues migrating, it transitions to yet another e-coherent state with $b_{coh1}$ discohesions. Unlike the $b_{coh2}$ and $b_{coh3}$ discohesions, $b_{coh1}$ discohesions take the form of small loops (blue diamond-shaped loops) with diameter of 22 Å and step height of $h = 0.34$ Å. Fig. 4(b) shows the discohesion line geometries for each case and the areal line densities as a function of interface position; the three transitions in e-coherent state are clearly visible. This simulation reveals that a single interface can have multiple possible e-coherent states, and that it may transition between states depending on the conditions. For the GB case here, these transitions are tied to GB motion.

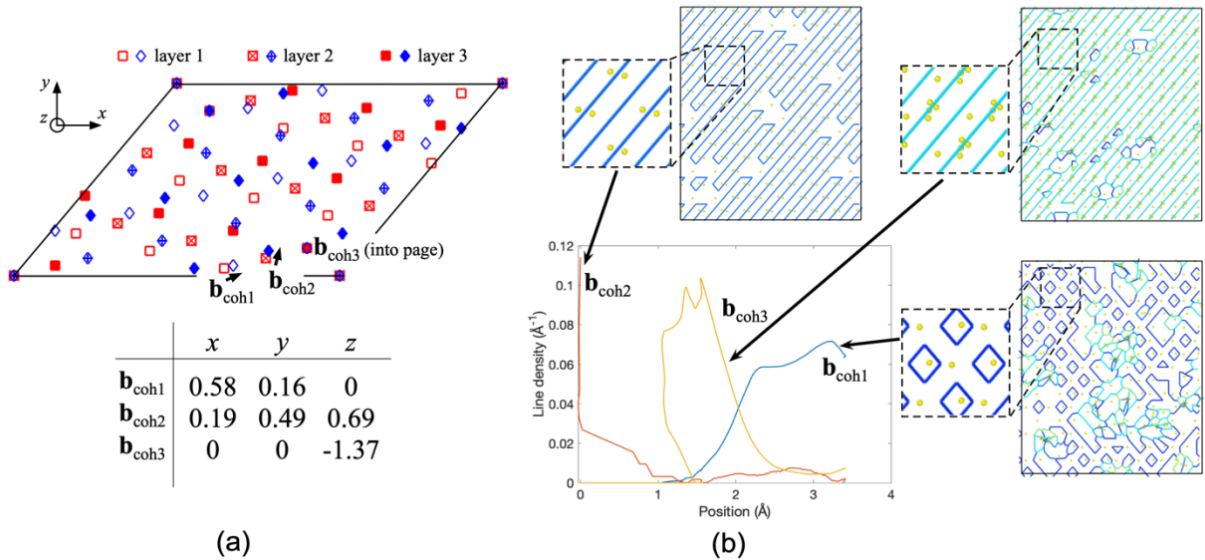

(a)   (b)

*Fig. 4 Migration of a e-coherent FCC GB. (a) Dichromatic pattern viewed orthogonal to the GB plane (with coincidence site lattice unit cell marked) showing the GB plane and atomic layers above and below it. (b) Transition of the e-coherent state during migration. Plot shows line density per unit area for the three different discohesions as a function of mean GB position. Snapshots show line defect structures dominated by the three discohesions, also showing CIS atoms in the interface. The table at the lower left gives the components of the Burgers vectors in units of Å.*

**Discussion**

E-coherency and discohesions provide an additional defect-based framework beyond that traditionally considered for characterizing the structure and dynamics of interfaces. The concepts of e-coherency and discohesions also complement existing concepts, such as the structural unit model and grain boundary complexions/phases, providing a richer language for describing and differentiating crystalline interfaces. Our simulations and comparisons with experiments show that the atomic structure of e-coherent interfaces is fundamentally different from traditional coherent interfaces, since both affine and non-affine CIS atoms are present. As a consequence, e-coherent interfaces always exhibit discohesion arrays in the relaxed interface. These defects are unrelated to semi-coherency or relaxation of lattice misfits or misorientation. Without the concepts of e-coherency and discohesions, there would be no way to explain the presence of these line defects or quantify their properties and geometry.

Our results also reveal how e-coherency can affect the behaviors and characteristics of an interface. The $\{1\,0\,\bar{1}\,1\}$ twin simulations demonstrate how e-coherency can affect interfacial dynamics, in this case by forcing the interface to move in two-plane increments. This was a direct consequence of the interface energy difference induced by the discohesion Burgers vectors at different interface plane positions. The $\alpha'$–$\varepsilon$ interface simulations (in the End Matter) revealed how e-coherency influences the topology of interfacial line defects. When other line defects, such as dislocations and disconnections are introduced into an e-coherent interface, they interact and react with the pre-existing discohesion content. In the α'–ε interface simulations, the different states of e-coherency at different plane positions led to two different disconnection modes co-existing within the same line defect loop. Such a behavior would never be observed in a traditional coherent interface, since it would lead to violation of conservation of Burgers vector.

In recent years the notion of GB complexions or phases has become increasingly studied and appreciated[16–18,35,36]. Each GB complexion/phase has a distinct atomic structure which dictates its associated properties, and often these complexions/phases are associated with different structural units[16,36]. GBs are able to transition from one complexion/phase to another via diffusive or displacive (e.g., dislocation) processes[18]. Our definition of e-coherency provides an additional concept of atomic structure variation for interfaces which is complementary to the concepts GB complexions/phases. For instance, some complexions/phases may be tied to a state of e-coherency, while others are not.

Another concept which is related to e-coherency is that of atomic shuffle (e.g., non-uniform atomic motions during interface migration)[22]. An e-coherent interface will always require atomic shuffle since the atomic motions associated with non-affine transformations are non-uniform (by definition). The major difference between the concepts of e-coherency and atomic shuffle is that e-coherency describes the state of the static interface, whereas shuffle describes atomic motions during migration. Hence, the two concepts synergize well to provide a more complete description of interfaces. This synergy is well illustrated by the example of the $\{1\,0\,\bar{1}\,1\}$ twin; previously it was argued that one-plane motion was unfavorable due to the associated large amount of shuffle which leads to slow migration kinetics[27]. However, our simulations show that one-plane motion does occur as an intermediate state on the way to forming a two-plane-high disconnection, demonstrating that one-plane motion is not kinetically slower than two-plane motion. This result provides evidence that it is the high energy of the one-plane step resulting from large Burgers vectors of the discohesions that prohibits one-plane motion, rather than the large shuffle distances.

The essential concept underlying e-coherency is that if a pair of atoms from the crystals are *close to coincidence* within the dichromatic pattern, the crystals may prefer to share a single CIS atom which is distorted away from both ideal lattice sites. In other words, we expect to see e-coherency in any interface where atom pairs in the dichromatic pattern/complex are relatively close together. This assertion is synonymous with stating that e-coherency will be preferred when the discohesion Burgers vector is small. Accordingly, e-coherency is more likely for interfaces where the affine CISs are spaced far apart (e.g., high-$\Sigma$ GBs) and with crystal structures having lower symmetry (e.g., HCP), since these conditions increase the likelihood that an atom pair will be nearly coincident. Our results in the text, End Matter, and the Supplementary Information identify e-coherent interfaces in FCC, BCC, and HCP GBs, as well as phase boundaries. Of course, the choice between forming a traditional coherent interface versus an e-coherent interface is dictated by the free energy of each structure, with the structure of lower free energy being the thermodynamically stable one.

Altogether, our experimental and simulation results show that concepts of e-coherency and discohesions have tangible physical manifestations that influence interface structure and properties. These results include interfacial line defect analysis of MD simulations based on rigorous identification of non-affine CIS atoms and one-to-one comparisons with HR-STEM. The description of e-coherency in terms of discohesion arrays provides a mathematically precise means for characterizing an e-coherent state while also giving insights into interface properties (e.g., energy). Based on our results and Supplemental Information, we anticipate that e-coherency is a commonplace feature of crystalline interfaces.

**Acknowledgements**

This work was supported by the U.S. Department of Energy, Office of Science, Basic Energy Sciences, Division of Materials Science and Engineering, under Award # DE-SC0022154 (T.J.M., S.H.C., and R.B.S.) and Award # FWP 18013170 (D.L.M. and A.H.). X.W.Z. gratefully acknowledges research support from the U.S. Department of Energy, Office of Energy Efficiency and Renewable Energy, Hydrogen and Fuel Cell Technologies Office through the H-Mat program. Sandia National Laboratories is a multi-mission laboratory managed and operated by National Technology & Engineering Solutions of Sandia, LLC (NTESS), a wholly owned subsidiary of Honeywell International Inc., for the U.S. Department of Energy's National Nuclear Security Administration (DOE/NNSA) under contract DE-NA0003525. This written work is authored, in part, by employees of NTESS. These employees, not NTESS, own the right, title, and interest in and to the written work and are responsible for its contents. Any subjective views or opinions that might be expressed in the written work do not necessarily represent the views of the U.S. Government. The publisher acknowledges that the U.S. Government retains a non-exclusive, paid-up, irrevocable, world-wide license to publish or reproduce the published form of this written work or allow others to do so, for U.S. Government purposes. The DOE will provide public access to results of federally sponsored research in accordance with the DOE Public Access Plan.

**Author contributions**

R.B.S contributed ideation, simulations, analysis, and writing. A.H. contributed microscopy, analysis, and editing. T.J.M. contributed simulations and analysis. S.H.C. contributed simulations and analysis. X.W.Z. contributed simulations, analysis, and editing. D.L.M. contributed ideation, microscopy, analysis, and writing.

**Data Availability**

Data supporting results presented in the manuscript are available at the following Data DOI: 10.5281/zenodo.14825899.

**Competing Interests**

None to declare.

# Supplementary Information for "E-coherent crystalline interfaces: coherency enhanced by discohesion arrays"

Ryan B. Sills[1*], Alejandro Hinojos[2], Trevor J. Murray[1], Shane H. Cooley[1], Xiaowang W. Zhou[2], and Douglas L. Medlin[2]

[1]Department of Materials Science and Engineering, Rutgers University, Piscataway, NJ 08854

[2]Sandia National Laboratories, Livermore, CA 94551

*Experiments*

The experimental, atomic resolution images presented in Figure 2(a) and Figure 3(a) of the main text were observed in deformed 304L austenitic stainless steel (SS) that had been charged with hydrogen prior to straining. The austenitic phase, γ (FCC), is metastable in 304L SS and transforms under deformation to ε-martensite (HCP) and to α'-martensite (BCC). The material in the present study was charged to 140 wppm hydrogen prior to straining under tensile load to 5% (true strain) at -50°C. Further details concerning this material, including its composition, hydrogen charging conditions, deformation procedures, and mechanical properties are provided in San Marchi et al.[1]. Thin foil disks for electron microscopic imaging were prepared by jet-electropolishing with an electrolyte of 10% perchloric acid and 90% ethanol using a Struers TenuPol 5 jet polisher operated at -11°C and 25.6 V. Imaging was conducted in a probe-corrected Thermo Fisher Scientific Themis Z Scanning Transmission Electron Microscope (STEM) operated at 300kV. The images were collected using a high angle annular dark field (HAADF) detector set to an angular collection range of 51-200 mrad. The electron probe was set to a semi-convergence angle of 25.2 mrad.

*corresponding author, ryan.sills@rutgers.edu

Figure S1 shows detail for the $(10\bar{1}1)$ HCP twin presented in Figure 2 of the main manuscript. This interface was observed at the intersection of two bands of ε-martensite indicated by the dark green box in Figure S1(a). Higher resolution detail of the intersection is provided in Figures S1(b) and (c). Figure S1(b) is from data collected at atomic resolution, but because the field of view is so large there is insufficient sampling in the figure to view the individually resolved atomic intensity peaks. We therefore also present a visualization of the data in Figure S1(c) by classifying the atomically resolved intensity peaks from Figure S1(b) into local atomic structural environments. This classification employed the 2-dimensional polyhedral template matching method (2D-PTM)[2] with an RMSD threshold value of $\tau = 0.05$ ($\tau$ is dimensionless and represents length normalized to the lattice parameter of the FCC phase). In Figure S1(c) the green and red regions are $\langle 1\,1\,0 \rangle$-oriented γ (FCC) and $\langle 2\,\bar{1}\,\bar{1}\,0 \rangle$-oriented ε (HCP), respectively. The analysis shows, for instance, that the ε-martensite bands are not perfect but instead possess faults corresponding to narrow strips of FCC (green) running parallel with the ε bands[3]. The atomic regions in the immediate vicinity of the twin are marked in blue. This is because near the interface the environments more closely match a two-dimensional projection of the $\langle 1\,1\,1 \rangle$-BCC structure (which was the third classification option for the 2D-PTM analysis) than they do the $\langle 1\,1\,0 \rangle$-oriented γ (FCC) and $\langle 2\,\bar{1}\,\bar{1}\,0 \rangle$-oriented ε (HCP). This is not to state that the local structure at the twin interface is in a BCC environment, only that of the three 2D-PTM classification options that we selected this was the best match. Additional peaks that failed to match either of the three options within a specified threshold are marked as black (for instance, there is a patch of poorly matched peaks at the right side of image for which the image quality was likely poor due to strain effects at the intersection point of the two ε-martensite bands). Figure S1(d), shows atomically resolved detail of the $(10\bar{1}1)$ twin. The image presented in Figure 2 of the main manuscript is taken from a section of this interface in a region between the faults in the ε-martensite.

As an additional comment, we note that each fault intersection is associated with an interfacial disconnection, two of which appear to possess single-plane high steps. Prior atomistic calculations of lattice dislocation interactions with $(10\bar{1}1)$ HCP twins[3] showed single-plane steps which were also associated with faults in the HCP. In that study, the faults were produced by emission of Shockley partial dislocations into the adjacent grain following decomposition of the lattice dislocation into the twin boundary. In our case, because the faults do not terminate within the field of view, we were not able to determine the full Burgers vector associated each disconnection and so were unable to assess whether an equivalent mechanism is at play at this specific interface.

Figure S2 provides background for the BCC-HCP interface presented in Figure 3 of the main manuscript. BCC α'-martensite often initiates at intersections with ε-martensite (HCP) bands[4,5]. Figure S2(a) shows an α'-nucleus (the bright central feature marked by the purple box) at the intersection of ε-martensite bands. Figures S2(b) and (c) show, respectively, atomic resolution HAADF-STEM image of the α' nucleus and the corresponding 2D-PTM analysis, again with an RMSD threshold of τ=0.05, of the local atomic environment, showing ⟨1 1 0⟩-oriented γ (FCC) (green), ⟨2 $\bar{1}$ $\bar{1}$ 0⟩-oriented ε (HCP) (red). and ⟨1 1 1⟩-oriented α' (BCC) (blue). Detail of the HAADF-STEM image and 2D-PTM analysis from the interfacial region near the top of the α'-nucleus are shown in Figures S2(d) and (e).

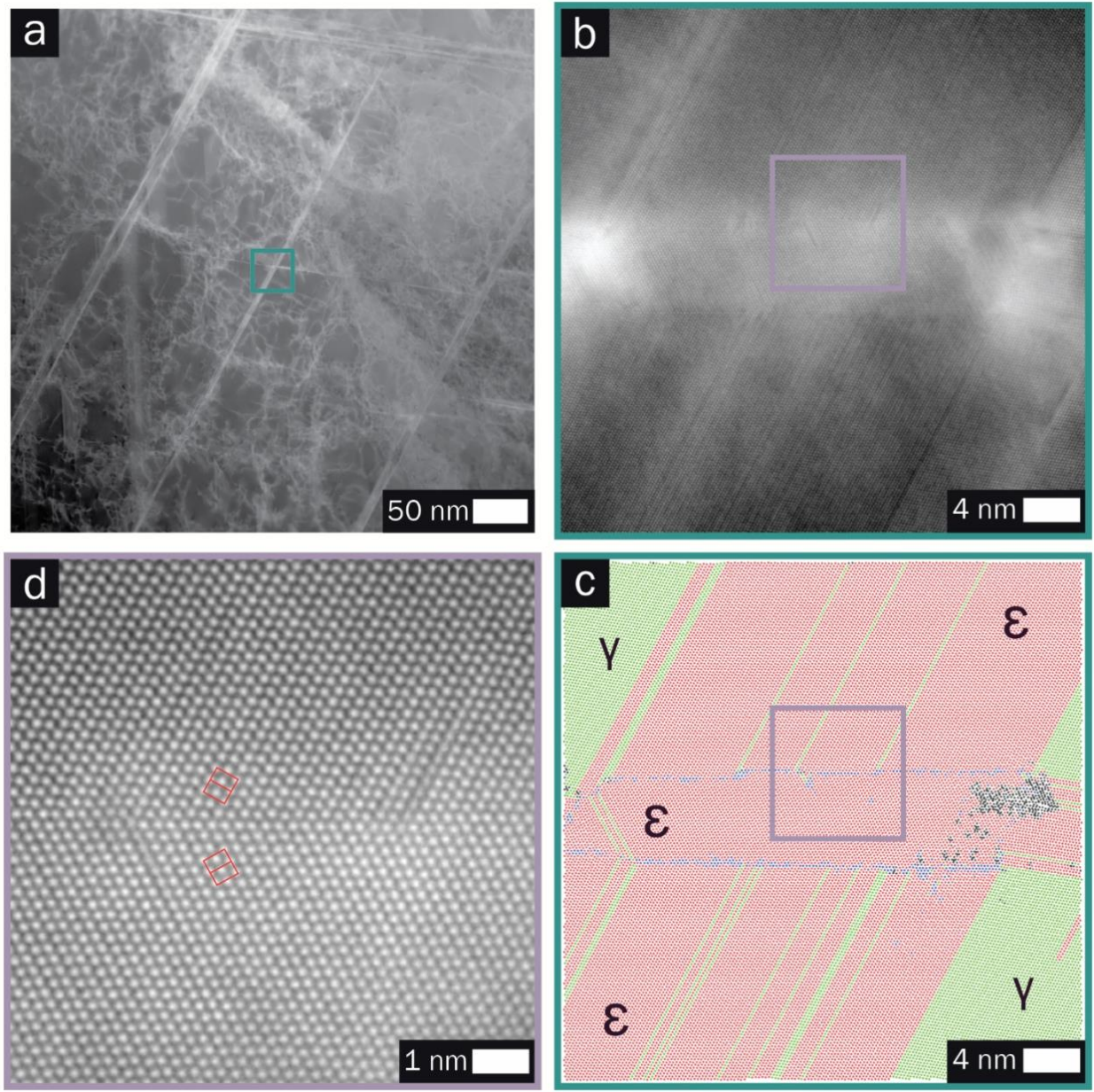

*Figure S1. Detail of ε-ε intersection in 304L stainless steel forming the $(10\bar{1}1)$ HCP twin discussed in Figure 2(a) of the main text. (a) Green box indicates the position of the intersecting ε-martensite bands. (b) HAADF-STEM image of the intersection (note: this image was collected at atomic resolution, but because the field of view is so large there is insufficient sampling in the figure to view the individual atomic columns). (c) Atomic intensity peak positions from (b)*

*classified into local atomic environments using the 2D-PTM method. γ-FCC ⟨110⟩ (green); ε-HCP ⟨2$\bar{1}\bar{1}$0⟩ (red). (d) Higher resolution detail of the (10$\bar{1}$1) twin (corresponding to area indicated by the purple box in (b) and (c)). HCP unit cells are indicated in the crystals on the two sides of interface.*

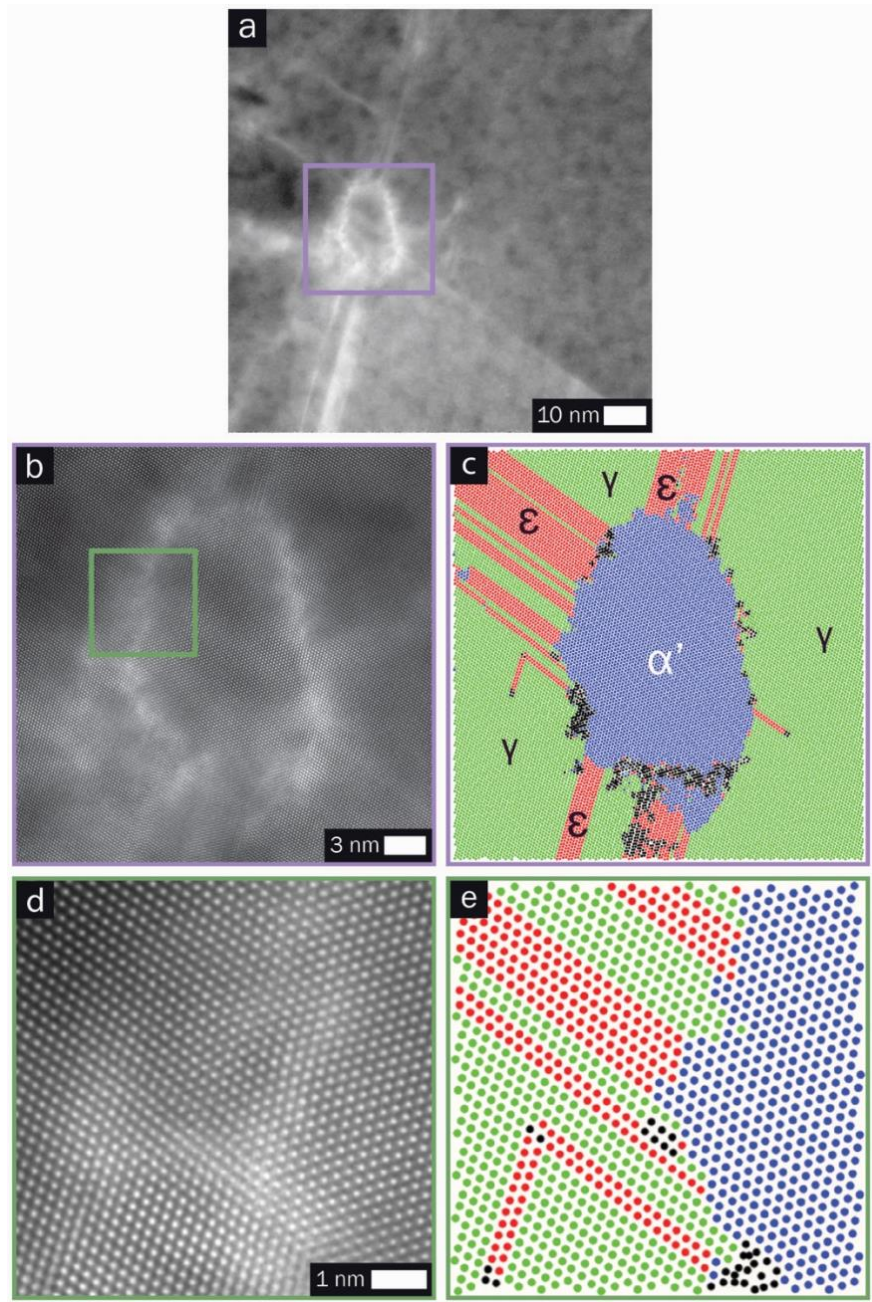

*Figure S2. (a) HAADF-STEM image of an α'-martensite (BCC) nucleus at the intersection of shear bands containing ε-martensite (HCP) in tensile-strained 304L stainless steel. High resolution HAADF image of the α' nucleus and a magnified image of the α'-ε interface are shown in Figure (b) and (d), respectively. (c) and (e) show the same areas as Figure (b) and (d), respectively, with the atomic peak positions colored with respect to their local atomic arrangement using the 2D-*

PTM method. γ-FCC ⟨110⟩ (green); ε-HCP ⟨2$\bar{1}\bar{1}$0⟩ (red); α'-BCC ⟨111⟩ (blue), and unidentified (black).

*Simulations*

*Methods*

All molecular dynamics (MD) simulations were performed in LAMMPS[6]. The twin and phase boundary simulations used the Zhou et al. embedded atom method potential for Fe[7] and the GB simulation used the Mishin et al. embedded atom method potential for Al[8]. Bicrystalline systems were constructed for the chosen crystal structures and orientations while enforcing periodic boundary conditions in the lateral (*x* and *y* directions). The *z*-axis was orthogonal to the interface plane in all cases, which was also periodic for the phase and GB simulations (a free surface condition was used for the twin simulation). Crystal orientations, box dimensions, and temperatures for the three case studies are given in Supplementary Table 1. When constructing the phase boundary, the BCC crystal was first inserted and the box shape was distorted to enforce coherency with the HCP crystal, which was then subsequently inserted. For all case studies, the system was first relaxed via energy minimization after which a dynamic simulation was run with an NPT ensemble setting system stresses to zero. The Nosé-Hoover thermostat and barostat used time constants of 0.01 and 1 ps, respectively. In the twin case study, shear strain was applied by applying a constant displacement rate in the *y*-direction to atoms at the top and bottom surfaces giving a shear strain rate of $10^8$ 1/s. With the GB test case, the eco synthetic driving force was used[9] (modified to allow for non-symmetric GBs) with an energy difference of 0.01 eV/Ω (where $\Omega = a^3/4$ is the atomic volume for the FCC crystals and *a* is the lattice constant) and a cut-off radius of 1.2*a*.

| | *Twin* | *Phase Boundary* | *GB* |
|---|---|---|---|

| $x_1$ | [2 $\bar{1}$ $\bar{1}$ 0] | [2 $\bar{1}$ $\bar{1}$ 0] | [7 6 $\bar{5}$] |
|---|---|---|---|
| $y_1$ | [0 1 $\bar{1}$ 2] | [0 1 $\bar{1}$ 2] | [9 $\bar{8}$ 3] |
| $z_1$ | implied | implied | [$\bar{2}$ $\overline{10}$ $\bar{6}$] |
| $x_2$ | [2 $\bar{1}$ $\bar{1}$ 0] | [$\bar{1}$ 1 1] | [9 5 2] |
| $y_2$ | [0 1 $\bar{1}$ $\bar{2}$] | [$\bar{1}$ 1 $\bar{2}$] | [1 3 $\overline{12}$] |
| $z_2$ | implied | [$\bar{1}$ $\bar{1}$ 0] | ($\bar{6}$ 10 2) |
| $L_x$ (Å) | 25.5 | 100.1 | 170.1 |
| $L_y$ (Å) | 28.4 | 141.5 | 201.2 |
| $L_z$ (Å) | 266.9 | 102.1 | 191.9 |
| $T$ (K) | 800 | 500 | 1 |

*Supplementary Table 1. Parameters for simulations presented in the manuscript.*

Line defect analysis used interfacial line defect analysis (ILDA) in OVITO Pro[10–12], which also uses polyhedral template matching (PTM) for phase identification[13]. The root mean square deviation (RMSD) parameter for PTM was set to 0.1, 0.078, and 0.08 Å for the twin, phase, and grain boundary case studies, respectively. These values were chosen based on the performance of polyhedral template matching when analyzing bulk crystalline atoms (e.g., to ensure that all bulk atoms were correctly identified); if the RMSD parameter is too small, then thermal fluctuations can cause misidentifications in the bulk. All coherent reference frames used with ILDA were manually specified based on the bicrystal geometries. Dichromatic patterns and associated line defects were analyzed using custom scripts in MATLAB.

*Additional Results*

We have observed two other e-coherent GBs which were not discussed in the main text and are now elaborated below. Supplementary Table 2 presents the crystal orientations and cell dimensions used in these studies. The first is a symmetric tilt FCC GB which has previously been studied extensively[14,15], the Σ5(3 1 0)[0 0 1] GB. We simulated this GB in Al using the Mishin et al.[8] potential at 300 K. A PTM RMSD value of 0.12 Å was used. The dichromatic pattern for this GB from a side view is shown in Fig. S3(a), which reveals a possible discohesion with Burgers vector $\mathbf{b}_{coh}$, separation distance $s = 5.57$ Å, and height $h = 0$ Å. The key feature is that two atoms are only 1.11 Å apart, making the structure unstable if both atoms are present. We construct the interface so that only one of these atoms is present in the system (by shifting the crystals), and after relaxing this system the resulting atomic structure and line defect content are shown in Fig. S3(b) and (c), respectively. The atomic structure follows the standard kite geometry that has been previously observed[14]. Here we reveal that this structure is associated with discohesion arrays, as shown in Fig. S3(c). The new coincidence sites introduced by the non-affine transformation are marked with "x" symbols.

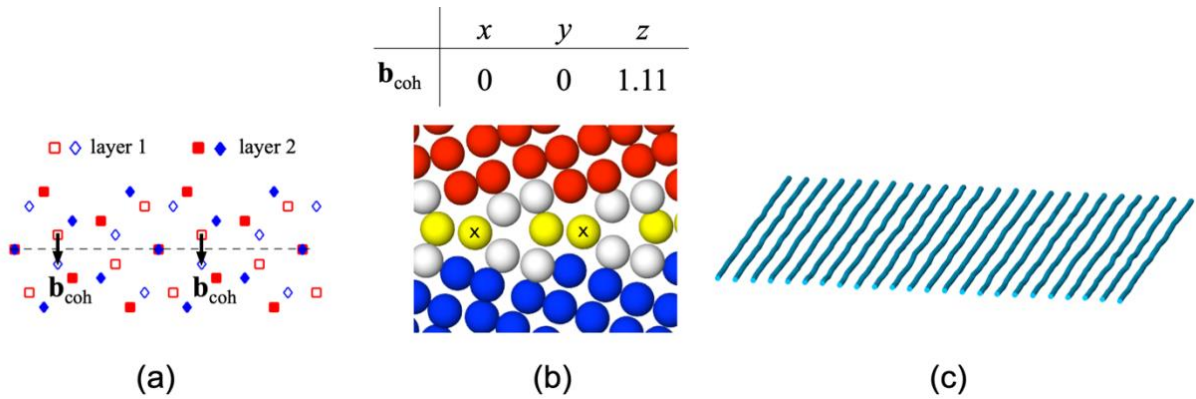

*Figure S3. E-coherent symmetric tilt Σ5(3 1 0)[0 0 1] FCC GB. (a) Side view of the dichromatic pattern of the GB with discohesions marked. (b) Atomic structure of the GB relaxed at 300 K, "x" symbols mark coincidence sites resulting from discohesion arrays. (c) Discohesion arrays present in the interface in (b).*

The other GB we have considered is a symmetric tilt $\Sigma 17(3\,3\,4)[1\,\bar{1}\,0]$ BCC GB which has previously been studied as well[16]. We simulate this GB in Fe at 300 K using the Zhou et al. potential[7]. A PTM RMSD value of 0.085 Å was used. The dichromatic pattern for the GB is shown in Fig. S4(a) and a possible discohesion with Burgers vector $\mathbf{b}_{coh1}$, spacing $s = 5.95$ Å, and height $h = 0$ is marked. Interestingly, if we first minimize the energy of the atomic structure we obtain linear discohesion arrays similar to the $\Sigma 5(310)$ FCC GB as shown in Fig. S4(b). However, unlike the FCC GB during an MD run the GB structure quickly becomes choppy and undulated, indicating that the pure e-coherent state is not very stable. Over time the GB is observed to fluctuate in terms of the defect content. The structure after 0.01 ns is shown in Fig. S4(c) which features small disconnection loops with Burgers vector $\mathbf{b}_1$ and step height $h = 1.24$ Å interspersed with discohesions, which are marked in the dichromatic pattern.

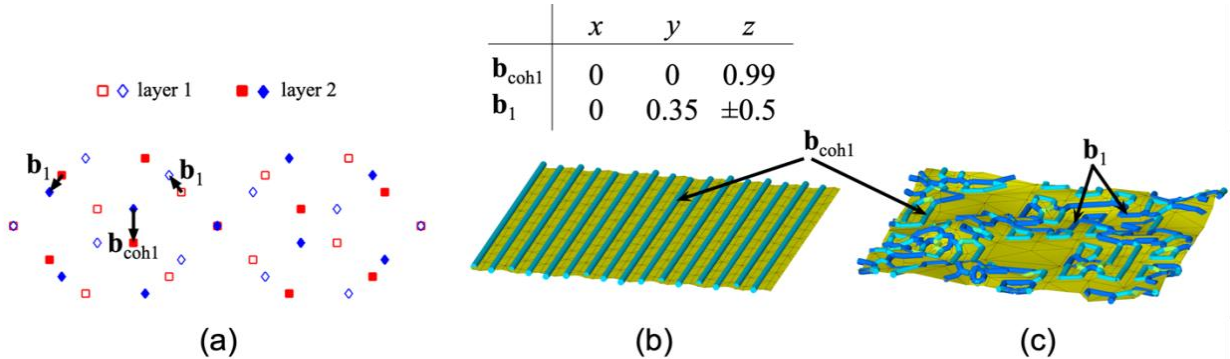

*Figure S4. E-coherent symmetric-tilt $\Sigma 17(3\,3\,4)[1\,\bar{1}\,0]$ BCC GB. (a) Dichromatic pattern of the side view of the GB with a discohesion and disconnection marked. (b) Initial state of the relaxed interface at low temperature (10 K) with discohesion arrays. (c) State of the interface at 300 K, showing a choppy and undulated structure with disconnection loops interspersed by discohesions.*

|        | Σ5 FCC GB | Σ17 BCC GB |
|--------|-----------|------------|
| $x_1$  | [0 0 1]   | [1 $\bar{1}$ 0] |
| $y_1$  | [1 3 0]   | [2 2 $\bar{3}$] |
| $z_1$  | [$\bar{3}$ 1 0] | [3 3 4] |
| $x_2$  | [0 0 1]   | [1 $\bar{1}$ 0] |
| $y_2$  | [$\bar{1}$ 3 0] | [$\bar{2}$ $\bar{2}$ $\bar{3}$] |
| $z_2$  | [$\bar{3}$ $\bar{1}$ 0] | [3 3 $\bar{4}$] |
| $L_x$ (Å) | 64.9    | 49.0       |
| $L_y$ (Å) | 76.9    | 47.6       |
| $L_z$ (Å) | 51.3    | 50.5       |
| $T$ (K)   | 300     | 300        |

*Supplementary Table 2. Parameters for simulations presented in the Supplementary Information.*

## *Diameter of a loop discohesion*

In the manuscript, we present an equation (Eq. (3)) for how to determine the spacing $s$ of straight discohesions. In some cases, discohesions form loop geometries instead (see manuscript Fig. 4(b)). Here we present how the diameter of a loop discohesion can be defined. Fig. S5 shows schematically how a loop discohesion can form with respect to the coherent interface, similar to Fig. 1 in the manuscript. Fig. S5 shows the atomic structure within and near the interface when viewed from above. The yellow atoms in the panel (a) are CIS atoms in the coherent interface. The blue and red atoms form pairs, one from each crystal, which are near to the interface plane and near to each other (but not coincident). By stretching each atom out of its perfect lattice position (and deleting one atom), these atoms can be rendered coincident leading to additional CIS atoms in the interface and the formation of an e-coherent interface (panel (b) of Fig. S5). Using the vectors, $^A t_i$ and $^B t_i$ defined in Fig. S5 (similar to the topological model[17]), the (average) diameter of the loop discohesion can be computed as follows:

$$d = \frac{2}{N_t}\sum_{i=1}^{N_t} \frac{\|^A\bar{t}_i + {^B\bar{t}_i}\|}{2} \tag{S1}$$

where $^I\bar{t}_i = {^I t_i} - ({^I t_i} \cdot \hat{n})\hat{n}$ is each vector orthogonalized with respect to the (coherent interface) terrace plane normal vector $\hat{n}$ and $N_t$ is the number of affine CIS atoms bounding the loop discohesion ($N_t = 4$ in Fig. S5).

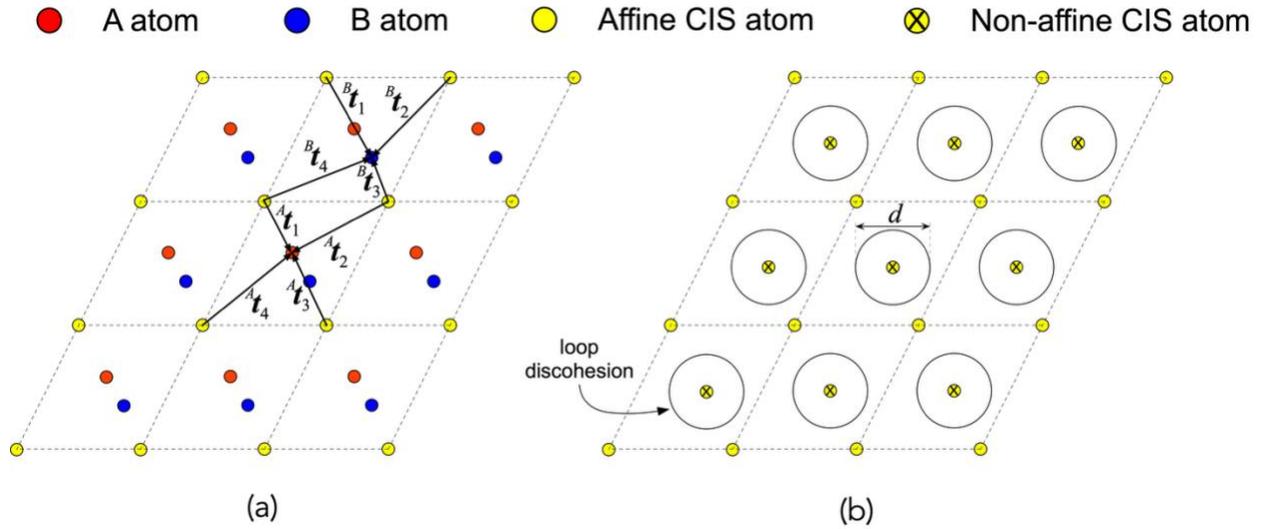

Figure S5. Schematic showing how to define the diameter d of a loop discohesion. (a) Coherent and (b) e-coherent atomic structures near the interface viewed orthogonal to the interface plane.

**Supplemental References**